\documentclass[10pt]{iopart}
\usepackage{epsfig,graphics,color}

\newcommand \lc {\langle}
\newcommand \rc {\rangle}

\newcommand \bvec{\left( \begin{array}{c} }
\newcommand \evec{\end{array} \right)}

\newcommand \bea{\begin{eqnarray} }
\newcommand \eea{\end{eqnarray} }

\newcommand {\be} {\begin{equation}}
\newcommand {\ee} {\end{equation}}

\begin{document}

\title[Jet Quenching]{Jet Quenching: the medium modification of the single and 
double fragmentation functions}

\author{A. Majumder}
\address{Nuclear Science Division, 
Lawrence Berkeley National Laboratory\\
1 Cyclotron road, Berkeley, CA 94720}

\date{ \today}

\begin{abstract}
The physics of the quenching of hard jets in dense matter is 
briefly reviewed. This is presented within the framework of the 
partonic medium modification of the fragmentation functions.  
Modifications in 
both deeply inelastic scattering (DIS) off large nuclei and
high-energy heavy-ion collisions are presented. 
\end{abstract}

\pacs{12.38.Mh, 11.10.Wx, 25.75.Dw}

%\preprint{LBNL-56705}

%\maketitle

The quenching of jets and the 
modification of jet structure has emerged as a new
diagnostic tool for the study of the partonic properties of dense
matter \cite{Gyulassy:2003mc}. The modification includes, not only a 
suppression of inclusive spectra of leading hadrons commonly 
referred to as jet quenching, but can be
extended to include the modification of many particle observables.
The simplest of these are the two-hadron correlations within the
jet cone. Such one and two particle observables have been measured
both in DIS \cite{dinezza04} and high-energy heavy-ion
collisions \cite{star,phenix}.

The single particle inclusive spectrum in DIS off a nucleus 
versus that in a nucleon (or deuterium) target 
demonstrates an increasing suppression as the momentum 
fraction $z$ of the detected hadron with respect to the 
initial partonic energy is increased. 
This suppression at 
a given $z$ has also been found to increase quadratically 
with the size of the nuclear target.  
For the two hadron correlation, a conditional ratio is 
measured: the distribution of associated hadrons given that there 
already exists a trigger hadron. This is measured 
in the case of a large nucleus 
and then divided by the same ratio in deuterium. 
The two-hadron correlation is, however, found to be very 
slightly suppressed, and almost independent of the nuclear 
size \cite{dinezza04}.  

In high energy $Au+Au$ collisions, the single inclusive 
spectra are suppressed compared to that in $p+p$ collisions.
What is presented is the differential cross section in $Au+Au$ collisions 
divided by the binary scaled differential cross section in $p+p$ collisions. 
This ratio is denoted as $R_{AA}$.  
Beyond a transverse momentum $p_T$ of 3~GeV for $\pi^0$'s and 6~GeV for charged hadrons, 
this suppression is 
almost a constant, independent of $p_T$ (at a $\sqrt{s} = 200$ GeV). 
The suppression is seen to increase with centrality with $R_{AA}$ 
assuming a value of barely one fifth for the most central events. 
In sharp contrast to this, the 
near side two particle correlation is moderately enhanced in central $Au+Au$
collisions relative to that in $p+p$ \cite{star,phenix}.

Theoretically, $n$-particle observables (where $n\geq 1$) 
in jet fragmentation
and their medium modification can be studied through $n$-hadron
fragmentation functions which can be defined as the expectation values 
of partonic field operators on $n$-hadron states. These
$n$-hadron fragmentation functions are non-perturbative and
involve long distance processes. However, they may be factorized
from the hard perturbative processes and their evolution with
momentum scale may be systematically studied in perturbative Quantum 
Chromo-Dynamics
(pQCD) \cite{amxnw}. 
Such an evolution is very similar to the well known case of $n=1$ 
hadron fragmentation functions \cite{col89}.

In these proceedings, the medium modification of
fragmentation functions in DIS off nuclei will be sketched within the
framework of generalized factorization and twist expansion \cite{guowang}.
The results are then extended to the case of
parton propagation
in heavy-ion collisions. 
Alternative methods for calculating the single inclusive distributions 
in heavy-ion collisions, 
based on a screened potential scattering model, have been investigated 
by many authors 
\cite{Gyulassy:1993hr,Baier:1996sk,Wiedemann:2000za}.
The advantage of the twist formalism \cite{guowang} lies, not merely in 
the ease of its applicability to DIS and heavy-ion collisions, but 
also in the presence of a single unknown normalization constant. 
This is set by fitting one experimental data point at one set 
of kinematic variables. 
With this parameter determined, one may predict the variation of 
the modified single hadron
fragmentation functions \cite{guowang,EW1}, as well as the medium
modification of two-hadron correlations. 
Such behaviour in both DIS off nuclei and
heavy-ion collisions are presented in comparison with experimental data.

Applying factorization to hadron production in single jet events in
DIS off a nucleus, $e(L_1)+A(p)\rightarrow e(L_2)+h_1(p_1)+ \ldots +X$,
one can obtain the $n$-hadron semi-inclusive cross section as,
\begin{equation}
E_{L_2}\frac{d\sigma^{h_1 \ldots}_{\rm DIS}}{d^3L_2 dz_1 \ldots}
=\frac{\alpha^2}{2\pi s}\frac{1}{Q^4}L_{\mu\nu}\frac{dW^{\mu\nu}}{dz_1 \ldots},
\end{equation}
where the ellipsis indicates the possible presence of more than one 
detected hadron. The semi-inclusive tensor at leading twist has the 
simple form,
\begin{eqnarray}
\frac{d W^{\mu \nu } }{d z_1 \ldots} &=&\sum_q \int d x  f^A_q(x,Q^2)
H^{\mu \nu}(x,p,q) D_q^{h_1 \ldots }(z_1, \ldots ,Q^2).
\end{eqnarray}
In the above, $D_q^{h_1...}(z_1...)$ is the $n$-hadron fragmentation
function,
$L_{\mu\nu}$ is the leptonic tensor,
the factor $H^{\mu \nu}$ represents the hard part of quark
scattering with a virtual photon which carries a four-momentum
$q=[-Q^2/2q^-,q^-,\vec{0}_\perp]$ and $f^A_q(x,Q^2)$ is the quark
distribution in the nucleus which has a total momentum
$A[p^+,0,\vec{0}_\perp]$. The hadron momentum fractions,
$z_n=p_n^-/q^-$ are defined with respect to
the initial momentum $q^-$ of the fragmenting quark.

\begin{figure}[htbp]
\hspace{0.5cm}
\resizebox{2in}{2in}{\includegraphics[1.25in,2.75in][6.75in,8.25in]{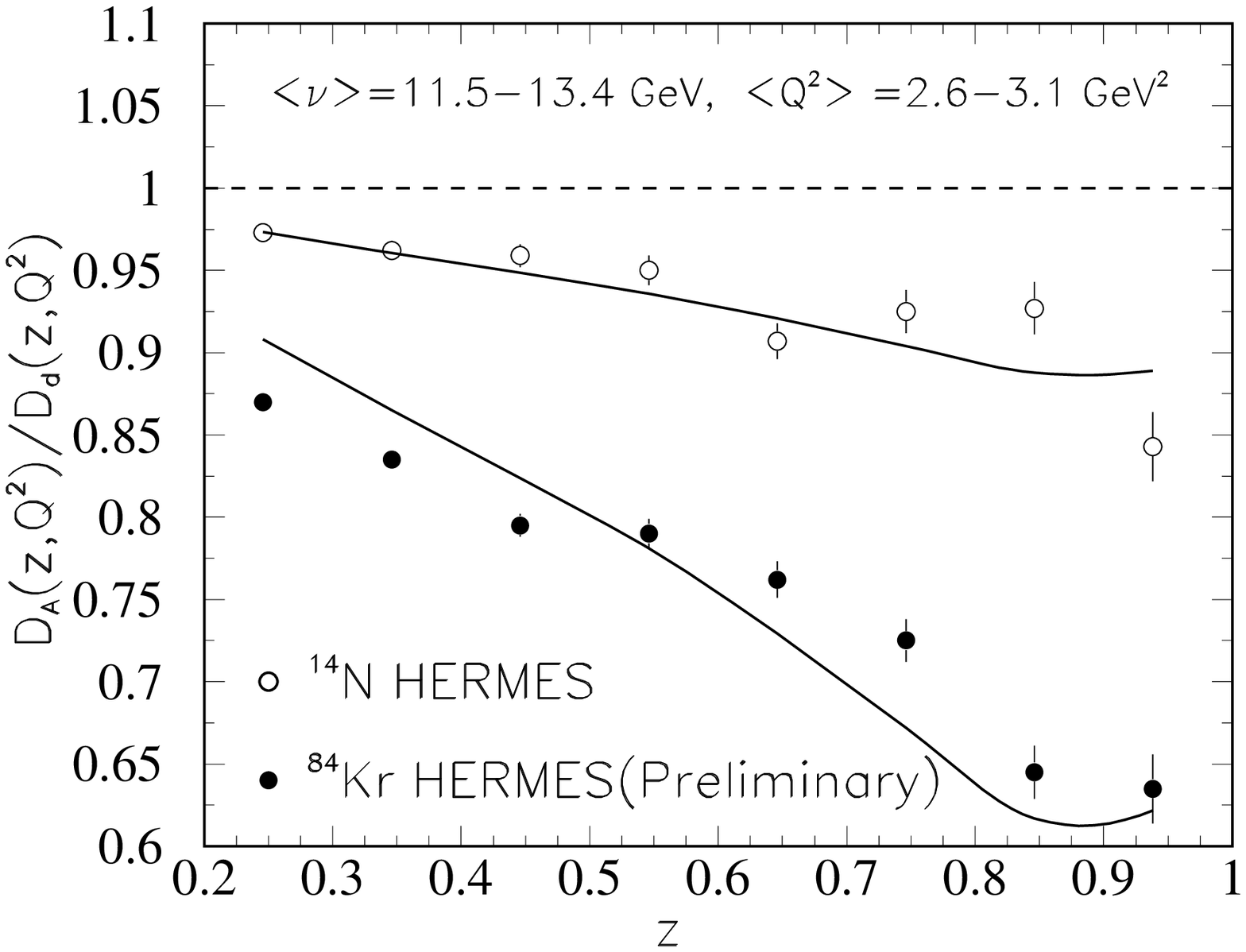}}
\hspace{1cm}
\resizebox{2in}{2in}{\includegraphics[0.5in,0.5in][6.0in,6.0in]{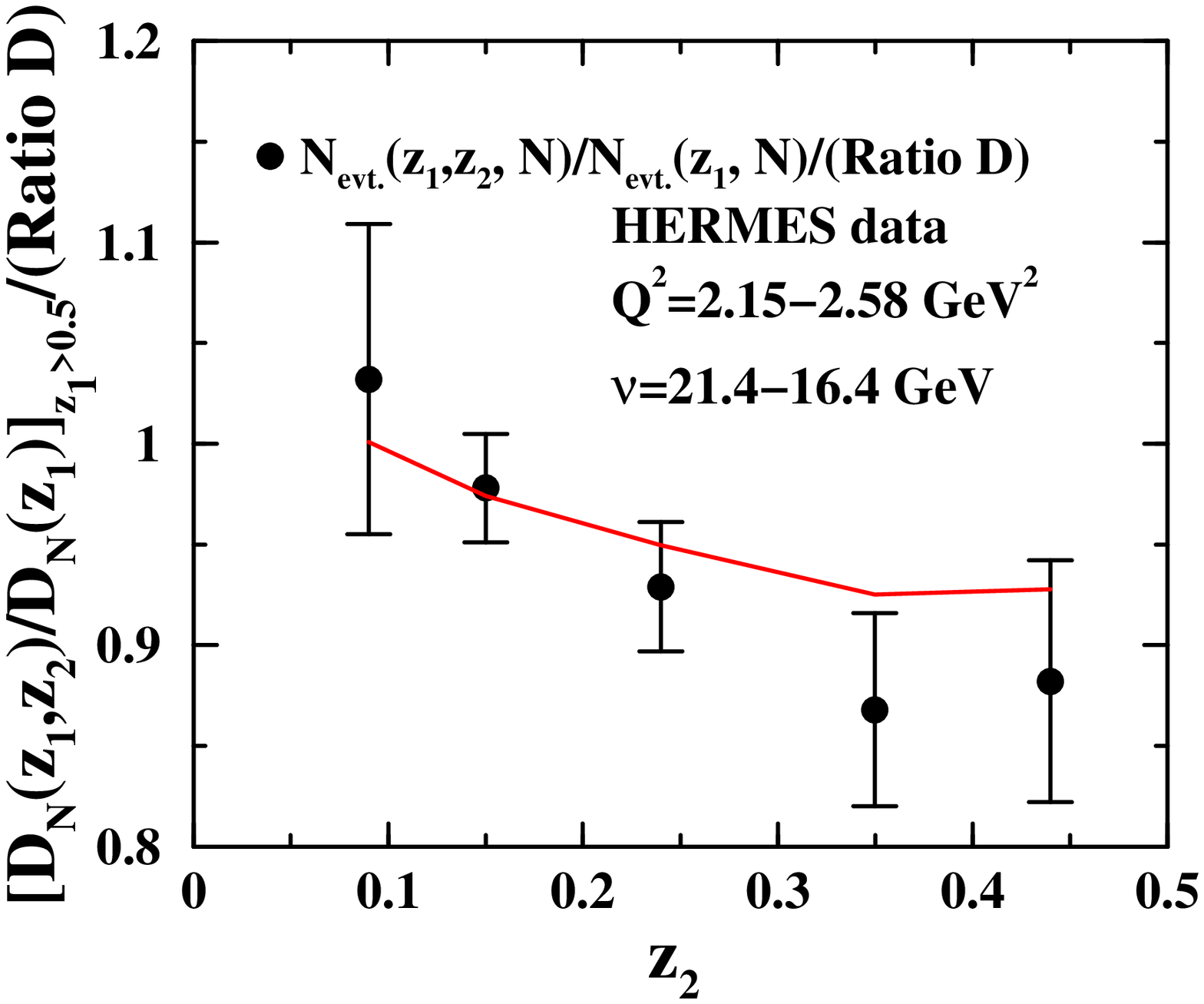}}
    \caption{ Results of the medium modification of
    the single (left plot, from Ref.~\cite{EW1}) and associated (right plot, from
    Ref.~\cite{maj04f}) hadron distribution, 
    in a cold nuclear medium
    versus its momentum fraction. 
    The ratio of the respective quantity in the nucleus 
    [Nitrogen (N) or Krypton (Kr)] with that in
    deuterium (D) are the HERMES data points. 
    In the plot for associated hadrons, the momentum fraction of the leading hadron
    $z_1$ is integrated over all allowed values above 0.5. }
    \label{fig1}
\end{figure}

At next-to-leading twist, the dihadron semi-inclusive tensor
receives contributions from multiple scattering of the struck
quark off soft gluons inside the nucleus with induced gluon
radiation. One can reorganize the total contribution
(leading and next-to-leading twist) into a product of the effective
quark distribution in a nucleus, the hard part of photon-quark
scattering $H^{\mu \nu}$ and a modified $n$-hadron fragmentation
function $\tilde{D}_q^{h_1 \dots } (z_1, \dots )$. The calculation of
the modified dihadron fragmentation function at 
next-to-leading twist in a nucleus \cite{maj04f} proceeds 
in similar fashion to that for the modified single hadron fragmentation
functions \cite{guowang}. The results of the medium modification of the 
single hadron fragmentation functions are shown in the left plot of Fig.~\ref{fig1}
in comparison with experimental data. The experimental point at the 
lowest $z$ in Nitrogen is used to fit the over all normalization constant. 
The variation with $z$ and with nuclear size is a prediction of the calculation and 
shows excellent agreement with the experimental results.

With no additional parameters, 
one can predict the nuclear modification of dihadron fragmentation
functions within the same kinematics. This is shown in the 
right hand plot of Fig.~\ref{fig1}, where results are presented 
for DIS off nitrogen.

%%%%%%%%%%%%%%%%%%%%%%%%%%%%%%%%%%%%%%%%%%%%%%%%%%%

In high-energy heavy-ion (or $p+p$ and $p+A$) collisions, jets are
always produced in back-to-back pairs. 
Correlations of two
high-$p_T$ hadrons in azimuthal angle generally have two Gaussian
peaks \cite{star,phenix}. Relative to the triggered hadron,
away-side hadrons come from the fragmentation of the away-side jet
and are related to single hadron fragmentation functions. On the
other hand, near-side hadrons come from the fragmentation of the
same jet as the triggered hadron and therefore the integral of the 
near side Gaussian peaks are related to dihadron fragmentation functions.

\begin{figure}[htbp]
\hspace{1cm}
  \resizebox{2in}{2in}{\includegraphics[2in,2.5in][8in,8.5in]{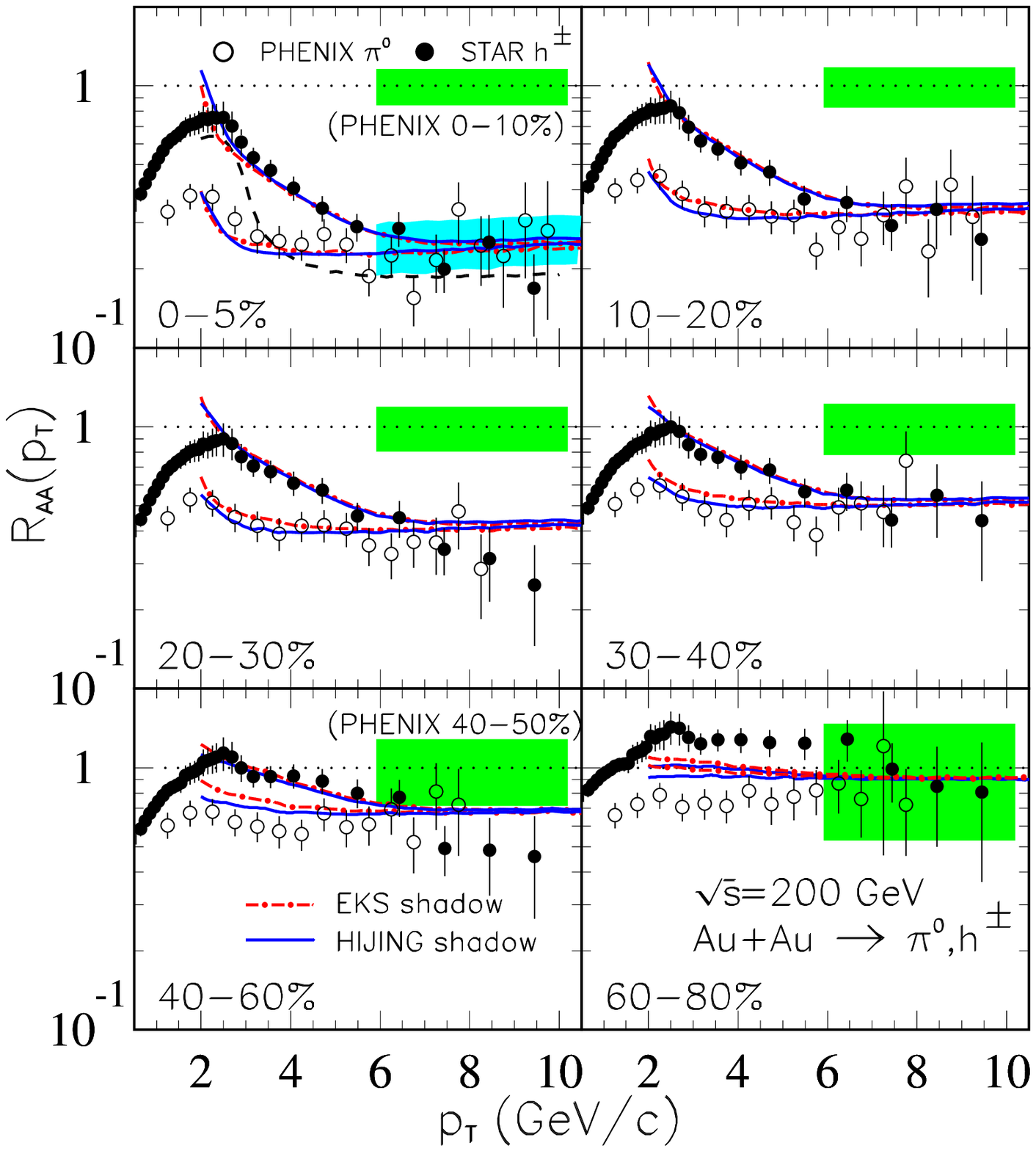}}
\hspace{1cm}
  \resizebox{2in}{2in}{\includegraphics[1.25in,0.5in][6.75in,6.0in]{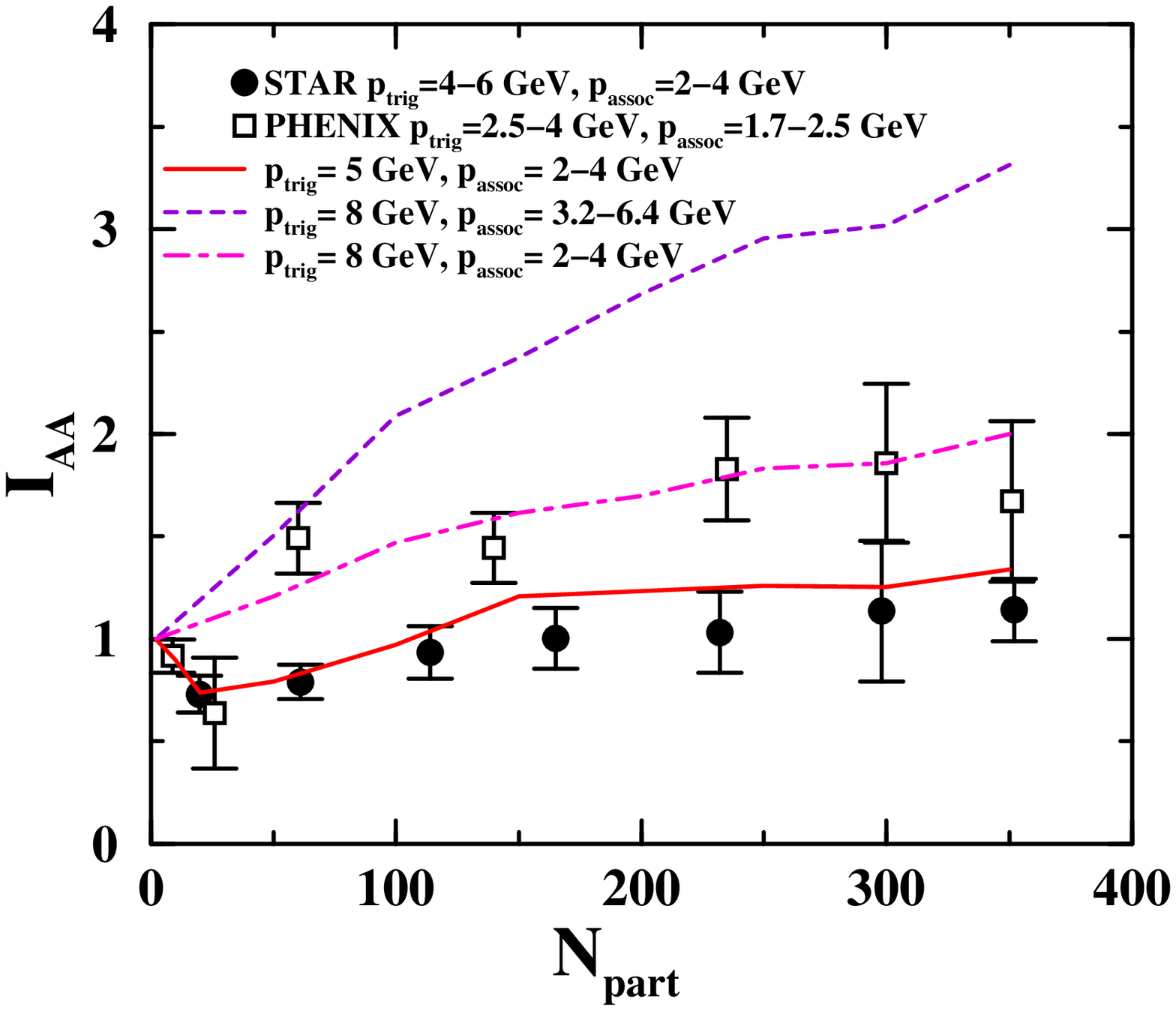}}
\caption{ Calculated medium modification of single (left plot, from Ref.~\cite{xnwang03}) 
          and associated (right plot, from Ref.~\cite{maj04f}) hadron
          distribution from jet fragmentation in $Au+Au$ collisions
          at $\sqrt{s}=200$ GeV. The left plot shows variation of $R_{AA}$ vs.
	  $p_T$ for different centralities. The right plot shows variation of 
	  $I_{AA}$ vs. centrality for different trigger and associated $p_T$ 
	  as compared to experimental data 
          \protect\cite{star,phenix}.} 
    \label{fig2}
\end{figure}

To extend the study of the medium modification of the fragmentation functions to
heavy-ion collisions, 
one also has to include
the effect of thermal gluon absorption \cite{EW2} such that the
effective energy dependence of the energy loss will be different
from the DIS case. Such a procedure, applied to the
study of the modification of the single hadron fragmentation functions 
successfully describes the quenching of single inclusive hadron
spectra, their azimuthal anisotropy and the suppression of the away-side
high $p_T$ hadron correlations \cite{xnwang03}. 
The result for the suppression of the single inclusive spectra 
as a function of the centrality of the collision is shown in the 
left side plot in Fig.~\ref{fig2}. The ratio $R_{AA}$ for $\pi^0$'s (PHENIX \cite{phenix}) 
and charged hadrons (STAR \cite{star}) is plotted.
The overall normalization constant is fitted at a given value of 
$p_T$ for the most central event. The variation of the suppression 
with $p_T$ and centrality is a prediction and shows very good agreement with the 
data. The over all gluon density is assumed to vary with centrality as 
the number of participants. Due to space constraints, we will skip the 
discussion of the suppression of the away side spectra and baryon meson differences. 
We refer the 
reader instead to Ref.~\cite{xnwang03} for this and further details regarding the figure.

The change of
the near-side correlation due to the
modification of dihadron fragmentation functions in heavy-ion
collisions can be similarly calculated. 
For a given value of $p_T^{\rm trig}$ of the
triggered hadron, one can calculate the average initial jet
energy $\lc E_T \rc$. Because of trigger bias and
parton energy loss, $\lc E_T \rc$ in heavy-ion collisions
is generally larger than that in $p+p$ collisions for a fixed
$p_T^{\rm trig}$ \cite{xnwang03}.

Using the overall parameter, 
determined in the single inclusive measurement, one calculates the modified dihadron
distributions. The ratio of such associated hadron distributions
in $Au+Au$ versus $p+p$ collisions, referred to as $I_{AA}$
\cite{star}, is plotted as the solid line in the right hand plot of Fig.~\ref{fig2}
together with the STAR data \cite{star}, as a function of the
number of participant nucleons. 
We also present data from PHENIX \cite{phenix} which 
accounts for correlations at a lower $p_T^{trig}$.
In central $Au+Au$ collisions, triggering on a high $p_T$ hadron 
biases toward a larger initial jet energy and therefore smaller 
$z_1$ and $z_2$. This leads to an enhancement in $I_{AA}$ due to
the shape of dihadron fragmentation functions \cite{amxnw}.
The enhancement increases with $N_{part}$ because of increased 
total energy loss. In the most peripheral collisions, the effect of
smaller energy loss is countered by the Cronin effect due
to initial state multiple scattering that biases toward smaller
$\lc E_T \rc$ relative to $p+p$ collisions. As a result, the
associated hadron distribution is slightly suppressed. 
For higher
$p_T^{\rm trig}$, the suppression will diminish because of the
disappearence of the Cronin effect (See Ref.~\cite{maj04f} for further discussions). 

\end{document}